\documentclass[doublecol]{epl2} 

\title{A Dirac type variant of the $xp$ model and the Riemann zeros}
\shorttitle{Title} 

\author{Kumar S. Gupta\inst{1} \and E. Harikumar\inst{2} \and Amilcar R. de Queiroz\inst{3}}

\institute{                    
  \inst{1} Saha Institute of Nuclear Physics, Theory Division, 1/AF Bidhannagar, Kolkata 700 064, India\\
  \inst{2} School of Physics, University of Hyderabad, Central University P O, Hyderabad, AP, PIN 500046, India\\
  \inst{3} Instituto de Fisica, Universidade de Brasilia, Caixa Postal 04455, 70919-970, Brasilia, DF, Brazil
}

\abstract{
We propose a Dirac type modification of the $xp$-model to a $x~\sigma\cdot p$ 
 model on a semi-infinite cylinder. This model is inspired by recent work of Sierra et al on the $xp$-model on the half-line. Our model realizes the Berry-Keating conjecture on the Riemann zeros. We indicate the connection of our model to that of gapped graphene with a supercritical Coulomb charge, which might provide a physical system for the study of the zeros of the Riemann Zeta function.}

\usepackage{latexsym}
\usepackage{amsmath}
\usepackage{amsfonts}
\usepackage{amssymb,slashed,dsfont}
\usepackage[latin1]{inputenc}
\usepackage{verbatim}

\begin{document}

\maketitle

There have been an increasing interest in physical models that might shed light 
on the Riemann hypothesis regarding the zeros of the Riemann Zeta functions. 
This program aims at a possible realization of a conjecture by Polya and 
Hilbert on the zeros of the Riemann Zeta function \cite{titchmarsh,edwards}. 
The Polya-Hilbert conjecture \cite{edwards} states that the complex zeros of 
the Riemann Zeta function on the critical line are described by the eigenvalues 
of a self-adjoint operator. One of the original proposals for such an operator 
arose from the work of Berry \cite{berry,berryk}, followed by Connes
\cite{connes} and Sierra \cite{sierra0,sierra1,sierra11,sierra12,sierra2,
sierra3}  with their collaborators. The attempts towards construction of such an operator are guided by various properties that are ascribed to it  \cite{berry,berryk}, including the requirements that it should exhibit a chaotic classical dynamics with isolated periodic trajectories and that it should break time reversal invariance. 

Guided by these requirements, Berry and Keating \cite{berryk} suggested that 
the operator $H_{0} = xp$ is a good candidate for the realization of the Polya-Hilbert  conjecture. The Berry-Keating operator $xp$ breaks time reversal invariance. However, it does not by itself leads to closed trajectories, which requires various additional identifications in the classical phase space. There are distinct identifications or regularizations leading to similar realizations of the Polya-Hilbert conjecture. However each such identifications also suggest distinct interpretations of the model at the level of a semi-classical analysis. There were two main distinct regularizations. One was proposed by Berry and Keating, which leads to a discrete spectrum for the model. Another regularization was proposed by Connes \cite{connes}, which leads to a continuum spectrum with a missing discrete set of eigenvalues. The spectrum counting in the case of Berry and Keating and the missing spectral line counting in the case of Connes are though similar. Both pictures were later seen to be equivalent by Sierra \cite{sierra11,sierra12} in a quantum model displaying a discrete spectrum inside a continuum.

An interesting physical realization of the Berry-Keating operator was given by 
Sierra and Townsend \cite{sierra1}, who observed that the $xp$-operator appears as the Hamiltonian of the quantum Hall effect when restricted to the lowest Landau level \cite{ezawa}. They further argued that from this perspective, the Berry-Keating Hamiltonian has a quantum description consistent with the Polya-Hilbert conjecture. The relationship of Riemann zeros to other quantum systems such as the inverted harmonic oscillator \cite{khare} and the Morse potential \cite{lagarias} have also been discussed in the literature.

Recently, Sierra and Rodriguez-Laguna \cite{sierra2} have proposed a new 
operator $H_1 = x \left ( p + \ell_p^2/p \right )$ where the particle 
is restricted to the (regularized) half-line $l_x \leq x \leq \infty$, with  $\ell_p^2$ a constant. The operator $H_1$ satisfies the essential properties of the Berry-Keating model and in addition gives rise to closed classical trajectories. In the quantum theory, the appearance of the $1/p$ term in $H_1$ leads  to a nonlocal boundary condition. With an appropriate choice of the kernel of $1/p$, $H_1$ was shown to admit a 1-parameter family of self-adjoint 
extensions with a spectrum that is bounded from below and is thus a legitimate 
quantum Hamiltonian operator. For certain choices of the self-adjoint extension 
parameter, this model seems to be consistent with the Polya-Hilbert conjecture.

The relation of the Hamiltonian operators in \cite{berry,berryk,connes,sierra1,
sierra2,sierra3} to the Polya-Hilbert conjecture appears through the Riemann-van Mangoldt formula
      \begin{equation}
       \label{Riemann-Mangoldt-1}
            N(E)=\langle N(E) \rangle + S(E),
      \end{equation}
where  $N(E)$ is the number of complex zeros of the Riemann Zeta function
with positive imaginary part less than $E$, $\langle N(E) \rangle$  is the
smooth part of $N(E)$ and $S(E)$ denotes a fluctuation in the counting (also
known as error function) given by
\begin{equation}
	\label{error-function-1}
      S(E)=\frac{1}{\pi}~\arg~\zeta\left(\frac{1}{2}+iE\right)=\mathcal{O}(\log(
E)).
\end{equation}
The smooth part of Eq. (\ref{Riemann-Mangoldt-1}) can be written as
\begin{equation}
	\label{smooth-part}
      \langle N(E) \rangle=\frac{\vartheta(E)}{\pi}+1,
\end{equation}
where the phase $\vartheta(E)$ of the Riemann Zeta function on the critical
line (also known as Riemann-Siegel function \cite{edwards}) is given by
\begin{equation}
\label{phasetheta}
      \vartheta(E)=\arg \Gamma\left(\frac{1}{4}+\frac{iE}{2} \right) -
\frac{1}{2} E \log\pi.
\end{equation}
Recall that by the use of Stirling's formula -- valid for large $E$ -- the above Riemann-Siegel function may be written as
\begin{equation}
  \vartheta(E)\sim \frac{E}{2}\log\frac{E}{2\pi}-\frac{E}{2}+\mathcal{O}(1).
\end{equation}

The Hamiltonian operators introduced in \cite{berry,berryk,connes,sierra1,
sierra2,sierra3} provide an estimate for the phase $\vartheta(E)$ of the 
Riemann Zeta function, which can be used in the Riemann-van Mangoldt formula
(\ref{Riemann-Mangoldt-1}) 
to make inferences about the zeros of the Riemann Zeta function.

In this Letter, following the spirit of the work of Sierra et al.
\cite{sierra2,sierra3}, we propose a variant of the $xp$ model on the half-line 
where the operator $p$ is replaced by a suitable two dimensional Dirac type 
operator $\slashed{p}$. The aim is to obtain a proper self-adjoint $xp$-operator
 on the half-line.

Let us consider the Dirac operator
\begin{equation}
      \slashed{p} \equiv  \sigma\cdot p =\sigma_x p_x + \sigma_y p_y,
\end{equation}
where $p_a=-i\partial_a$, with $a=x,y$, and $\sigma_x,\sigma_y$ are Pauli 
matrices. Thus,
\begin{equation}
      \slashed{p} = -i  \left( \begin{array}{cc}
                              0 & \partial_x -i \partial_y \\
                              \partial_x+i\partial_y & 0
                        \end{array} \right).
\end{equation}
We assume that this operator acts on two-component column-vectors valued on a Hilbert space defined on a semi-infinite cylinder. The semi-infinite cylinder is described by
$0\leq x < \infty$ and the $y$-direction is a circle of radius $R$. There are a 
$U(1)$ worth of possible boundary conditions that guarantees conservation of 
probability. This statement is equivalent to say that there is a $U(1)$ family 
of self-adjoint operators $\slashed{p}$ on the semi-infinite cylinder.

The above considerations on the operator $\slashed{p}$  should be contrasted 
with the fact that there is no self-adjoint operator $p$ on the half-line. This 
statement can be checked by computing the deficiency indexes $n_\pm$ and then 
use the criteria discovered by Von Neumann \cite{rs,essin2006}. The 
deficiency indexes $n_\pm$ are defined as the dimension of the space of square-integrable solutions of the equations $p \psi_\pm=\pm i \psi_\pm$. For the operator $p$ on the half line, we have $n_+=1$ and $n_-=0$. Thus, by von Neumann theorem, there is no 
self-adjoint operator $p$ on the half-line. Equivalently, there is no boundary 
condition on half-line that guarantees conservation of probability.

We now consider the Hamiltonian $H=x\slashed{p}$. This is our proposed 
modified version of the $xp$-model. This Hamiltonian acts on a suitable domain 
of the Hilbert space of square-integrable functions on the cylinder. Let us 
consider the normal ordered operator\footnote{In the final stages of this work, 
we learned that a very similar model to this one was already suggested by M. 
Asorey, J. Esteve and G. Sierra on unpublished notes. Their motivations was the 
same as ours.}
\begin{equation}
\label{Dirac-xp-model-1}
      H=\sqrt{x}~\slashed{p}~\sqrt{x}.
\end{equation}
Similar to the operator $\slashed{p}$, there is a $U(1)$ family of self-adjoint 
operators $H$ on the semi-infinite cylinder. Indeed, $H$ inherits the same 
possible self-adjoint domains from $\slashed{p}$. We observe that a 
different choice of normal ordering in the definition of $H$, like for instance 
$H=(1/2)(x\slashed{p}+\slashed{p}x)$ does not change the conclusions of the 
following analysis.

Observe that the Hamiltonian (\ref{Dirac-xp-model-1}) is time-reversal odd. In the present case, time-reversal transformations are performed by $\Theta= \exp(i\pi\sigma_y/2)K$, with $K$ being a complex conjugate operator, such that it is an anti-unitary transformation $\Theta^2=-\mathds{1}$. It is straightforward to see that $\Theta H \Theta^{-1}=-H$. This fact means that for appropriate boundary conditions the spectrum of the quantized Hamiltonian contains time conjugate pairs, that is, eigenfunctions with energy $E$ is mapped to eigenfunctions with energy $-E$ under $\Theta$. This will play a crucial role later in our analysis.

For a two-component column-vector
\begin{equation}
      \Psi(x,y)=\left( \begin{array}{c}
                               \psi_1 \\
                               \psi_2
                        \end{array} \right),
\end{equation}
the eigenvalue equation $H\Psi=E\Psi$ leads to
\begin{align}
      \label{Dirac-like-eq-1}
     \sqrt{x} \left(\partial_x -i \partial_y \right)\sqrt{x} \psi_2 &= iE \psi_1 ,
\\
      \sqrt{x}\left( \partial_x +i\partial_y \right) \sqrt{x} \psi_1 &= iE \psi_2 .\label{Dirac-like-eq-2}
\end{align}
This system of equation is equivalent to a second order differential equation 
for one of the components, let us say $\psi_1(x,y)$. From the topology of the 
$y$-direction, we may consider the \emph{Ansatz}
\begin{align}
 \label{Ansatz-11}
      \psi_1(x,y) &= \frac{1}{\sqrt{2\pi}}~\frac{\varphi (x)}{x}~ 
e^{i\frac{n+\alpha}{R}y} , \\
       \psi_2 (x,y) &= \frac{1}{\sqrt{2\pi}}~\frac{\chi (x)}{x}~ 
e^{i\frac{n+\alpha}{R}y} , \label{Ansatz-12}
\end{align}
with $n\in\mathbb{Z}$ and $\alpha\in [0,1)$. The integer $n$ is associated with the winding number due to the compactification of $y$-direction.

The analysis presented below is valid for each value of $n$ and we assume it to fixed to some arbitrary value without loss of generality. The parameter $\alpha$ is associated with the phase of the quasi-periodic boundary conditions along $y$-direction. It parameterizes the family of self-adjoint domains of the Hamiltonian. In more physical terms, $\alpha$ is related to a flux passing through the circle obtained by the compactification of the
$y$-direction. If we now set 
	\begin{equation}
		\label{x-u-n-alpha-1}
      x= \frac{R u}{2 (n + \alpha)}, 
	\end{equation}
we obtain
\begin{equation}
      \label{whitt}
      \frac{ \partial^2 \varphi}{\partial u^2}+\left[ -\frac{1}{4} + \frac{1}{2
u} + \frac{E_{n}^2+\frac{1}{4}}{u^2} \right]~\varphi = 0,
\end{equation}
which is exactly of the Whittaker's form \cite{AS}. It may be noted that the 
expression for $\psi_2$ may be obtained from that of $\psi_1$.

The time-reversal invariance aspects of the present model are the following. The Dirac operators in equations (\ref{Dirac-like-eq-1}) and (\ref{Dirac-like-eq-2}) transform correctly under time reversal. Observe that time-reversal transformation flips the two components of the spinor and changes the sign of the energy. However the ansatz (\ref{Ansatz-11}) and (\ref{Ansatz-12}) violate time-reversal symmetry. The reason for this is the following. Note first that the parameter $\alpha$ which characterizes the domain of self-adjointness of the momentum operator supported on the compact $y$ direction has the physical interpretation of a magnetic flux passing through the circle. The magnetic field automatically breaks time reversal invariance. Now, even if $\alpha=0$, the axis of the cylinder is semi-infinite and the corresponding coordinate can always be taken as positive. Thus even for $\alpha=0$, the quantity $x$ or $u$ in Eq. (\ref{x-u-n-alpha-1}) must be taken as positive. This means that $n$ is a positive integer. This in turn implies that the winding modes are chiral. Hence the time reversal invariance is broken. 

In order to set up the allowed boundary conditions, we first regularize the 
semi-infinite line $0\leq x < \infty$ to $x_0 \leq x < \infty$, for a positive 
$x_0$. Equivalently we can say that 
\begin{equation}
\label{cutoff}
u_0 = \frac{2}{R}( n + \alpha ) x_0
\end{equation}
At $x=x_0$ or equivalently at $u = u_0$ we consider the boundary condition
      \begin{equation}
	    \label{bound-cond-1}
            \varphi(u_0)=0.
      \end{equation}
Note that if both $n=0$ and $\alpha = 0$, then $u_0 = 0$ even if $x_0 \neq 0$. 
In what follows, we shall assume that values of $n$ and $\alpha$ are such that 
$(n + \alpha) \neq 0$. Later we discuss the case for small but finite $u_0$.

The analysis of boundary conditions presented below can be cast in a more formal language of self-adjoint domains and the von Neumann theorem \cite{rs,essin2006}. The boundary condition on the wave-functions for the singular attractive inverse-square potential following from the self-adjoint extension has been discussed in \cite{voronov,case}. The close relation between 
the boundary conditions following from the introduction of a cut-off to that 
obtained from self-adjoint extension has also been discussed in the literature \cite{essin2006,case,jackiw}. Here we prefer to report on this route of regularization and renormalization, since it is more familiar among physicists.

The general solution of Whittaker equation with boundary condition
(\ref{bound-cond-1}) is (apart of an overall multiplicative constant)
\begin{align}
      \label{solution-gen-1}
      \varphi(u)&=\left[ M_{\frac{1}{2},-iE_{n}}(u_0)~M_{\frac{1}{2},+iE_{n}}(u) \right. \nonumber \\
      &\left.~-M_{\frac{1}{2},+iE_{n}}(u_0)~M_{\frac{1}{2},-iE_{n}}(u)\right],
\end{align}
where $M_{k,m}(x)$ is Whittaker function (see equation 13.1.32 of \cite{AS}.)
\begin{align}
      \label{whittaker-def-1}
      M_{k,m}(u) = e^{-\frac{u}{2}}u^{m+\frac{1}{2}}M\left(m-k+\frac{1}{2}, 
1+2m ;u\right), 
\end{align}
with $M(a,b;u)$ being Kummer's confluent hypergeometric functions \cite{AS}. 

We observe, before proceeding to the analysis of the spectrum itself, that the solution $\varphi(u)$ in (\ref{solution-gen-1}) is odd under time-reversal. Indeed, by mapping $E_n$ to $-E_n$, $\varphi(u)$ goes to $-\varphi(u)$, as it should.

We would like to have a solution $\varphi(u)$ that is square-integrable. For that we first note that as $u \rightarrow \infty$ \cite{AS}, 
\begin{equation*}
      M(a,b;u)\approx \frac{e^{i\pi a}
\Gamma(b)}{\Gamma(b-a)}~u^{-a}+\frac{\Gamma(b)}{\Gamma(a)}~e^{u}~u^{a-b}
+\mathcal{O}\left(\frac{1}{u}\right).
\end{equation*}
Therefore, as $u \rightarrow \infty$,
\begin{align}
\varphi(u) &\rightarrow A e^{\frac{u}{2}} \frac{1}{\sqrt{u}} \left[
      M_{\frac{1}{2}, -iE_n} (u_0) \frac{\Gamma(1 + 2iE_n)}{\Gamma(+iE_n)} \right. \nonumber \\
      &\left. ~ -M_{\frac{1}{2}, +iE_n} (u_0) \frac{\Gamma(1 - 2iE_n)}{\Gamma(-iE_n)} 
\right]
\end{align}
Thus the requirement of square-integrability implies the condition 
\begin{equation}
\label{transc-eq-1}
\frac{M_{\frac{1}{2},-iE_n}(u_0)}{M_{\frac{1}{2},+iE_n}(u_0)} 
= \frac{\Gamma(1 - 2iE_n)}{\Gamma(1 + 2iE_n)}  \frac{\Gamma(+iE_n)}{\Gamma(-iE_n)} .
\end{equation}

We now analyze condition (\ref{transc-eq-1}) in the limit $u_0 \rightarrow 0$, but finite, which corresponds to a small but finite cutoff. Observe that ideally one would like to take $u_0$ equal zero. However, due to quantum instability associated with a strongly attractive inverse square potential, we cannot really remove it. Equivalently, if $u_0$ were zero, there would be a continuous  spectrum of energy. Yet, it is sensible to consider an expansion for small $u_0$. In this case, by using certain properties of the Gamma function (6.1.32 of \cite{AS}) and of the Whittaker function (\ref{whittaker-def-1}), we obtain\footnote{The small $u_0$ expansion of the R.H.S. of (\ref{transc-eq-1}) goes like
\begin{equation*}
  \frac{M_{\frac{1}{2},-iE_n}(u_0)}{M_{\frac{1}{2},+iE_n}(u_0)} = u_0^{-2i E_n}\left( 1- \frac{2iE_n~u_0}{1+4E_n^2}+\mathcal{O}\left(u_0^2\right) \right).
\end{equation*}
} 
\begin{equation}
	\label{main-condition}
  \frac{\Gamma\left(\frac{1}{4}+\frac{iE_n}{2} \right)^2}{\Gamma\left(\frac{1}{4}-\frac{iE_n}{2} \right)^2} \left(\frac{u_0}{8} \right)^{-2iE_n}
=  \cot\left(\frac{\pi}{4}+i\frac{\pi E_n}{2}\right)
\end{equation}

We now consider the limit of asymptotically large values of $E_n$. In this asymptotic limit, using Eqs. (\ref{main-condition}) and (\ref{phasetheta}), we obtain
\begin{equation}
\label{sim1}
e^{i 4 \vartheta (E_n)}  \left(\frac{u_0}{8 \pi} \right)^{-i2E_n} e^{- i \pi} = 1
\end{equation}
This condition (\ref{sim1}) is comparable (but not the same), and indeed has the same spirit, as Eq. (22) of \cite{sierra1} (instead of the dimensionless parameter $L^2/\ell^2$ there, here we have the parameter $u_0$). 

After we take the logarithm of both sides of Eq. (\ref{sim1}) and use (\ref{smooth-part}), we obtain
\begin{equation}
\label{finalresult}
2 \left [ \frac{E_n}{2 \pi} \log \left (\frac{u_0}{8 \pi} \right )  + \frac{9}{8} \right ] - 2 \langle N(E_n)\rangle = N_{E_n},
\end{equation}
where $N_{E_n}$ is an integer, which, following \cite{sierra1}, can be interpreted as counting the number of states whose energy is less than some fixed $E_n$. Note that the factor of 2 on the l.h.s. of Eq. (\ref{finalresult}) indicates that in this model the number of missing spectral lines is twice as much as in the usual $xp$ model. The Eq. (\ref{finalresult}) is almost identical to Eq. (23) of \cite{sierra1}.

It may be noted that the total energy of the system has a degeneracy arising from the quantity $n$. However, as pointed out before, the above analysis is valid for each fixed value of $n$. In addition, the interesting results arise from the analysis of the equation in the semi-infinite direction of the cylinder. Thus, the degeneracy arising from the appearance of the quantity $n$ can be removed by fixing it to some arbitrary value, without loss of generality.

We have thus far shown that the $x\slashed{p}$ model together with the boundary 
condition (\ref{bound-cond-1}) leads to an algebraic equation for the spectrum
 given by (\ref{main-condition}). On one side, this algebraic condition contains 
the phase of the Riemann-Zeta function (\ref{phasetheta}) which is part of the 
Riemann-van Mangoldt counting formula (\ref{Riemann-Mangoldt-1}). On the other 
side, apart from the ratios of certain Gamma functions, it contains an undetermined parameter, namely the cut-off $u_0$. 

The appearance of the cut-off $u_0$ in (\ref{bound-cond-1}) deserves some 
comments. As mentioned earlier, the cutoff $u_0$ is related to the self-adjoint extension parameter of the strongly attractive 
inverse square potential on the half-line \cite{case,voronov}. Similar to the discussions in \cite{essin2006,jackiw} of a conformal 
quantum mechanics, this cut-off implies a breaking of the scale symmetry by 
quantization. That 
means a scale anomaly emerges in this problem. Indeed, from a classical perspective, 
the $xp$ model, and also the $x\slashed{p}$, is scale invariant. Also from 
another perspective, the Hamiltonian associated with Eq. (\ref{whitt}) in the 
short distance limit is singular and scale invariant. But the boundary condition
 (\ref{bound-cond-1}) breaks this scale invariance. The analysis of such 
singular behavior for equations like Eq. (\ref{whitt}) is given in \cite{case}. 
The works \cite{aguado,esteve,camblong} also contain  important discussions for 
the emergence of scale anomaly in a similar context.

The restoration of the scale anomaly at the quantum level is relevant for the 
problem of Riemann Zeta function. Indeed, that allows us to enforce some of the 
requirements of Berry and Keating at the quantum level. One mechanism for the 
restoration of an otherwise anomalous symmetry is by the use of an appropriate 
mixed state \cite{queiroz}. This restoration mechanism in respect to the present problem is under current investigation and will be reported elsewhere.

Another objective of this Letter is to present, following the spirit of
the work of Sierra and Townsend \cite{sierra1}, a possible physical realizations 
for the $H=x\slashed{p}$ model. 

To that end, let us first consider the Calogero model, whose solutions can be 
classified using the degree $k$ of a 
polynomial that appears in the analysis \cite{calogero}. It was shown in
\cite{guptaghosh} that for $k=0$, the Calogero model with a complex coupling has
real 
eigenvalues. The corresponding differential equation (see Eq. (12) of
\cite{guptaghosh}) has exactly the same form as Eq. (\ref{whitt}), although the 
parameters appearing in the equations in these two cases have different
interpretations. The main similarity between the two lies in the fact that they
both 
describe a system of strongly coupled inverse square interaction. Such a system 
was first discussed by Landau, who observed that the strongly attractive 
inverse square 
coupling leads to a quantum instability that was characterized by ``fall to the 
centre'' \cite{landau}. Subsequently, a more complete quantum treatment was 
given by Case \cite{case} who found that the spectrum is unbounded from below
and is characterized by a one-parameter family of an undetermined constant. In
technical terms, this unknown constant is nothing but the self-
adjoint extension parameter, which encodes in itself the combined effects of 
the short distance physics. In this work, as well as in \cite{guptaghosh}, the 
same role is played by the cut-off $u_0$. Hence, the essential physics described 
here is that of a Calogero type system with a strongly attractive inverse 
square interaction.

This naturally leads to the question if there is indeed a physical system which 
realizes such a potential. If a physical system exists whose eigenvalue equation
 is exactly governed by Eq. (\ref{whitt}), then the spectrum of that physical 
system would provide a concrete realization of the Polya-Hilbert conjecture. 
While we do not have an exact answer to this question, we notice that there is a graphene system whose equations are again very similar to what has been described above. 

It is now well known that dynamics of low energy excitations in graphene is 
governed by a Dirac equation \cite{sem}. For our purpose, we consider the case 
of gapped graphene \cite{ho,novi}. In the presence of an external supercritical 
Coulomb impurity \cite{novi,castro,levi1,levi2}, the effective radial equation for gapped graphene 
\cite{sen1} has the same formal structure as that of Eq. (\ref{whitt}), although
 again with a different interpretation of the parameters. In addition, the 
solutions of a gapped graphene system with supercritical Coulomb charge
\cite{sen1} have the same form as Eq. (\ref{whitt}) described above. When the
external  Coulomb charge in graphene is strongly attractive or equivalently in
the  supercritical region, the system again exhibits a quantum instability and
``fall to the centre''  \cite{castro,levi1,levi2,sen1,sen2,sen3}, whose
qualitative behaviour is similar to what we find here. In a graphene system,
such an instability manifests itself through characteristic features in the
local density of states (LDOS), which in principle can be measured using
scanning tunneling microscope (STM) \cite{castro,levi1,levi2}. With 
suitable reinterpretation of the system parameters, the STM measurements of the 
LDOS in gapped graphene with a supercritical Coulomb charge may yield valuable 
information regarding the spectrum of our model considered here.

It may also be mentioned that the same equation obtained in \cite{guptaghosh}, 
appears in the analysis of non-relativistic AdS/CFT correspondence \cite{moroz}.
 Typically, such strongly attractive inverse square operators arise also in the 
analysis of near-horizon conformal structure of black holes \cite{ksg1,ksg2}. 
While none of these problems are identical to what has been considered here, 
the structural similarity of the equations and the corresponding solutions 
perhaps indicates a deeper relationship between these systems.

The systems described above share a common feature that they all correspond to 
a strongly attractive inverse square potential characterized by quantum
instability. In such systems, it is possible to employ a renormalization group
scheme to address the issue of the instability \cite{rajeev}. The instability
indicates an incomplete understanding of the short distance physics. Typically
such a system can be analyzed by first introducing a cut-off and then studying
the renormalization group flow of the relevant parameters of the system, given
by the corresponding $\beta$-function. It would be interesting to study the 
consequences of the RG flow for the system described here.

In summary, in this Letter, we have proposed a Dirac type variant of the $H=xp$ 
given in Eq. (\ref{Dirac-xp-model-1}). We then analysed its spectrum and 
related it with the Riemann-van Mangoldt counting formula for the complex zeros 
of zeta function \cite{edwards}. An important observation is that in order to 
find this quantum  spectrum the scale invariance of the classical $xp$-model is 
necessarily broken. Based on this fact, we believe that the restoration of the 
scale invariance at the quantum level will have crucial implications to the 
Hilbert-Polya conjecture. Finally we suggested a physical realization of the 
present model in terms of the dynamics of the low energy excitations of the 
graphene in the presence of an external supercritical Coulomb impurity.

\emph{Acknowledgments}: The authors would like to thank Prof. A. P. Balachandran
 and Prof. A. Reyes for discussions that led to this work. We would also like 
to thank Prof. M. Asorey, Prof. J. Esteve and Prof. G. Sierra for reading and 
commenting on a preliminary version of this work. ARQ also thanks Prof. B. 
Carneiro Cunha and Prof. H. Nazareno for discussions at the final stage of the 
work. ARQ acknowledges the warm hospitality of  T.R. Govindarajan at The 
Institute of Mathematical Sciences, Chennai, where the main part of this work 
was done. ARQ is supported by CNPq under process number 307760/2009-0.



\begin{thebibliography}{99}
 
\bibitem{titchmarsh} E. C. Titchmarsh, {\it The Theory of the Riemann Zeta-
Function} (Oxford University Press, Oxford, 1951)

\bibitem{edwards} H.M. Edwards, {\it Riemann's Zeta Function} (Academic Press, 
New York, 1974).

\bibitem{berry} M. V. Berry in {\it Quantum Chaos and Statistical Nuclear 
Physics}, ed. T. H. Seligman and H. Nishioka, Springer Lecture Notes in Physics 
Vol. 263 (Springer, New York, 1986).

\bibitem{berryk} M. V. Berry and J. P. Keating, SIAM Rev. {\bf 41}, 236 (1999).

\bibitem{connes} A. Connes, Selecta Mathematica, New Series {\bf 5}, 29 (1999).

\bibitem{sierra0} G. Sierra, Nucl. Phys., {\bf B776}, 327-364, (2007).

\bibitem{sierra1} G. Sierra and P. K. Townsend, Phys. Rev. Lett. {\bf 101}, 
110201 (2008).

\bibitem{sierra11} G. Sierra, New J. Phys., {\bf 10}, 033016, (2008).

\bibitem{sierra12} G. Sierra, J. Phys., {\bf A41}, 304041, (2008).

\bibitem{sierra2} G. Sierra and J. Rodriguez-Laguna, Phys. Rev. Lett. {\bf 106},
 200201 (2011).

\bibitem{sierra3} G. Sierra,  J. Phys.  {\bf A45}, 055209 (2012).

\bibitem{ezawa} Z. F. Ezawa, {\it Quantum Hall Effect}, World Scientific Publishing Company (2008).

\bibitem{khare} R. K. Bhaduri, Avinash Khare, S. M. Riemann and E. L. Tomusiak, Ann. Phys. {\bf 254}, 25 (1997).

\bibitem{lagarias} J. C. Lagarias, Communications in Number Theory and Physics {\bf 3}, 323 (2009), arxiv:0712.3238[math.SP].

\bibitem{rs} Reed and Simon, {\it Methods of Modern Mathematical Physics} Academic Press, New York (1978).

\bibitem{essin2006} A. M. Essin and D. J. Griffiths, Am. J. Phys., {\bf 74}, 109-117, (2006).

\bibitem{AS} M. Abromowitz and I. A. Stegun, {\it Handbook of Mathematical 
Functions}, Dover, New York, 1974.



\bibitem{voronov} D.M. Gitman, I.V. Tyutin and B.L. Voronov, arXiv:0903.5277 (quant-ph).

\bibitem{case} K. M. Case, Phys. Rev. {\bf 80}, 797 (1950).


\bibitem{jackiw} R. Jackiw, Beq Memorial Volume, edited by A. Ali and P. 
Hoodbhoy (World Scientific, Singapore, 1991).

\bibitem{aguado} Aguado, M., Asorey, M. and Esteve, J.  Commun. Math. Phys., 
Springer, 2001, 218, 233-244

\bibitem{esteve} J. G. Esteve, Phys. Rev. D, {\bf 66}, 125013, (2002).

\bibitem{camblong} Camblong, H. E. and  Ordonez, C. R., Phys. Rev. D, American 
Physical Society, 2003, 68, 125013

\bibitem{queiroz} Balachandran, A. P. and Queiroz, A. R., Phys. Rev., 2012, D85,
 025017.


\bibitem{calogero} F. Calogero, J. Math. Phys. {\bf 12}, 419 (1971). 

\bibitem{guptaghosh} Pijush K. Ghosh and Kumar S. Gupta, Phys. Lett. {\bf A 323}, 29 (2004).

\bibitem{landau} L.D.Landau and E. M. Lifshitz, {\it Quantum Mechanics}, 
Butterworth-Heinemann, (1981).


\bibitem{sem} G. W. Semenoff, Phys. Rev. Lett. {\bf 53}, 2449 (1984).

\bibitem{ho} V. R. Khalilov and C. L. Ho, Mod. Phys. Lett. {\bf A 13}, 615 (
1998).

\bibitem{novi} D. S. Novikov, Phys. Rev. {\bf B 76}, 245435 (2007).

\bibitem{castro} V. M. Pereira, J. Nilsson and A. H. Castro Neto, Phys. Rev. 
Lett. {\bf 99}, 166802 (2007).

\bibitem{levi1} A. V. Shytov, M. I. Katsnelson and L. S. Levitov, Phys. Rev. 
Lett. {\bf 99}, 236801 (2007).

\bibitem{levi2} A. V. Shytov, M. I. Katsnelson and L. S. Levitov, Phys. Rev. 
Lett. {\bf 99}, 246802 (2007).

\bibitem{sen1} Kumar S. Gupta and Siddhartha Sen,  Phys.Rev. {\bf B78}, 205429 (
2008).

\bibitem{sen2} Kumar S. Gupta and Siddhartha Sen, Mod. Phys. Lett. {\bf A24}, 
99 (2009).

\bibitem{sen3} Baishali Chakrabarti, Kumar S. Gupta and Siddhartha Sen, Phys.
Rev. {\bf B83}, 115412 (2011).

\bibitem{moroz} S. Moroz, Phys. Rev. {\bf D81}, 066002 (2010).

\bibitem{ksg1} Kumar S. Gupta and Siddhartha Sen, Phys. Lett. {\bf B505}, 191 (
2001).

\bibitem{ksg2} Kumar S. Gupta and Siddhartha Sen,  Phys. Lett. {\bf B526}, 121 (
2002).

\bibitem{rajeev} Kumar S. Gupta and S. G. Rajeev, Phys. Rev. {\bf D48}, 5940 (
1993).




\end{thebibliography}

\end{document}